\documentclass[doublecol]{epl2}
\usepackage{amsmath}
\title{First order magneto-structural transition and magnetocaloric effect in MnNiGe$_{0.9}$Ga$_{0.1}$}

\author{Pallab Bag \and R Nath}
\institute{School of Physics, Indian Institute of Science Education and Research Thiruvananthapuram, Kerala-695006, India}
\date{\today}

\abstract{The first order magneto-structural transition ($T_t\simeq95$~K) and magnetocaloric effect in MnNiGe$_{0.9}$Ga$_{0.1}$ are studied via powder x-ray diffraction and magnetization measurements. Temperature dependent x-ray diffraction measurements reveal that the magneto-structural transition remains incomplete down to 23~K, resulting in a coexistence of antiferromagnetic and ferromagnetic phases at low temperatures. The fraction of the high temperature Ni$_2$In-type hexagonal ferromagnetic and low temperature TiNiSi-type orthorhombic antiferromagnetic phases is estimated to be $\sim 40\%$ and $\sim 60\%$, respectively at 23~K. The ferromagnetic phase fraction increases with increasing field which is found to be in non-equilibrium state and gives rise to a weak re-entrant transition while warming under field-cooled condition. It shows a large inverse magnetocaloric effect across the magneto-structural transition and a conventional magnetocaloric effect across the second order paramagnetic to ferromagnetic transition. The relative cooling power which characterises the performance of a magnetic refrigerant material is found to be reasonably high compared to the other reported magnetocaloric alloys.}
\begin{document}
\maketitle
\section{Introduction}
Heusler alloys with magneto-structural transition (MST) are receiving increasing attention in current condensed matter physics due to their fundamental and technological relevance. In such materials, manifestation of MST often gives rise to interesting properties such as magnetocaloric effect (MCE)\cite{Liu2012NatureCom,Krenke2005Nature},magnetoresistance\cite{Yu2006APL,Sharma2006APL}, field induced shape memory/strain effect\cite{Kainuma2006Nature,Ullakko1996APL}, glass like magnetic states\cite{Sharma2007PRB,Banerjee2011PRB} etc. In particular, magnetic refrigeration based on the MCE has been thought as a possible alternative for the vapour compression refrigeration technique, though it is far from being possible. Therefore, technological advances demand new and improved materials with MST at elevated temperatures.

The compound MnNiGe belong to the class of half Heusler alloys. It undergoes a structural transition from high temperature Ni$_2$In-type hexagonal with space group $P6_3/mmc$ (hex) to low temperature TiNiSi-type orthorhombic with space group $Pnma$ (orth) phase  at $T_{\rm t} \sim 470$~K and the low temperature phase shows the onset of a spiral antiferromagnetic (AFM) ordering at $T_{\rm N} \simeq 350$~K\cite{Bazela1976PSSA,Anzai1978PRB,Fjellvaag1985JMMM}. The first order structural transition ($T_{\rm t}$) can be tuned over a wide temperature range by chemical substitution either at the magnetic Mn/Ni site by Fe/Co/Cr\cite{Aryal2015JAP,Liu2013APL,Liu2012NatureCom,zhang2010JPDAP,Quetz2014JAP} or at the non-magnetic Ge site by Al/Ga/Sn\cite{Shi2013PSSA,Samanta2012APL,Liu2011IEEE}. It is predicted that when $T_{\rm t}$ is reduced below the Curie temperature $\theta_{\rm p}$ of Ni$_2$In-type hex phase [below which it is ferromagnetic (FM)], the structural transition gets coupled inductively with the FM state of Ni$_2$In-type hex phase resulting in a first order MST at $T_t$\cite{Zhang2008APL,zhang2010JPDAP,Liu2012NatureCom}. At the MST, the transition may occurs either from low temperature orth AFM to high temperature hex FM phase \cite{Quetz2013JAP,Zhang2013JAP,Aryal2015JAP,zhang2010JPDAP} or from low temperature orth FM to high temperature hex paramagnetic (PM) phase\cite{Samanta2012APL,Liu2012NatureCom}.

The Ga substitution at the Ge site (MnNiGe$_{1-x}$Ga$_x$) leads to a MST from low temperature orth AFM to high temperature hex FM phase and the MST varies from $\sim 200$~K to $\sim 140$~K with increasing Ga concentration upto 8.5\%\cite{Zhang2013JAP,Shi2013PSSA,Dutta2016JPDAP}. It has been noticed that in almost all the compounds in this series, the magnetization at low fields, well below the MST remains non-zero in contrast to the zero value anticipated in an AFM state. Zhang et al\cite{Zhang2013JAP} and Dutta et al\cite{Dutta2016JPDAP} predicted that when the sample is cooled below MST, the structural transformation is not complete and a small amount of FM hex phase exists with the major AFM orth phase at low temperatures. However, this coexistence of phases is not yet detected using direct experiments in any of the compounds of this series. In this work, we first time report the 10\% Ga doped MnNiGe$_{0.9}$Ga$_{0.1}$ alloy and investigate the magneto-structural phase transition as well as MCE and quantitatively estimate the coexisting hex FM and orth AFM phase fractions below $T_t$.
\section{Experimental details}
Polycrystalline MnNiGe$_{0.9}$Ga$_{0.1}$ sample was prepared by arc melting the constituent elements, weighed in the desired stoichiometry. The ingot was remelted several times to get better homogeneity and then annealed at 850~$^0$C continuously for five days. The weight loss of the sample was found to be $\sim 0.8$\%. Phase purity of the sample was confirmed from the powder x-ray diffraction (XRD) measurements (PANalytical powder diffractometer with Cu K$_\alpha$ radiation) carried out as a function of temperature $T$ (23~K$\leq T \leq 300$~K) using a low temperature attachment (Oxford Phenix). Magnetization ($M$) was measured as a function of temperature (2~K$\leq T \leq 380$~K) and magnetic field $H$ (-9~T$\leq H \leq 9$~T) using a vibrating sample magnetometer (VSM) attachment to the physical property measurement system (PPMS, Quantum Design).
\section{Results and discussion}
\begin{figure}[t!]
	\centering
	\includegraphics[width=\linewidth,height=.9\linewidth]{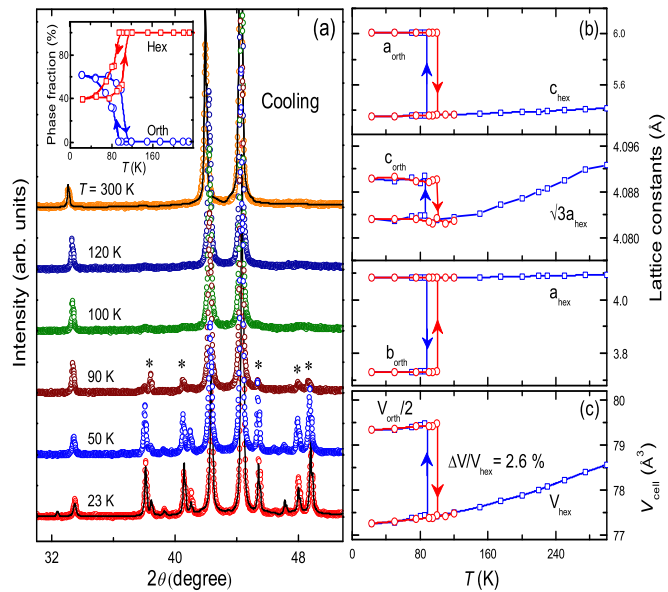}
	\caption{(a) Temperature dependent powder XRD measured while cooling from 300~K to 23~K. At $T=90$~K, the peaks corresponding to orth phase are marked by asterisk. Inset: the phase fractions (\%) of the hex FM and orth AFM phases vs. $T$. (b) Lattice constants and (c) unit cell volume ($V_{\rm cell}$) vs. $T$ for cooling and heating protocols (indicated by arrows).}
	\label{Fig1}
\end{figure}

In order to study the structural transition in MnNiGe$_{0.9}$Ga$_{0.1}$, we performed powder XRD as a function of temperature during cooling from 300~K to 23~K and subsequent heating. Figure~\ref{Fig1}(a) shows some representative XRD pattern collected at different temperatures during cooling only. At 300~K, all the Bragg peaks could be indexed based on the Ni$_2$In-type hex phase\cite{Anzai1978PRB,Fjellvaag1985JMMM}. At $T \simeq 90$~K, some extra peaks were noticed in addition to the peaks of the hex phase which were identified to be due to the TiNiSi-type orth phase. This indicates a structural transition from high temperature Ni$_2$In-type hex phase to low temperature TiNiSi-type orth phase at $T^{*}\simeq90$~K during cooling. However, during heating, the structural transition appears at a slightly higher temperature $T^{**}\simeq100$~K reflecting a thermal hysteresis. Here, we take the structural transition temperature $T_t\simeq95$~K as the average of $T^{*}$ and $T^{**}$. With reducing temperature ($T\leq T_t$), the intensity of the peaks corresponding to the orth phase increases while it decreases for the high temperature hex phase. This implies that the structural transition at $T_t \simeq 95$~K is incomplete and both orth and hex phases coexist below $T_{\rm t}$.

To estimate the lattice parameters and the phase fractions, Rietveld refinement was performed on the XRD data using \verb FullProf  package\cite{Rodriguez1993PhysicaB}. The initial structural parameters for this purpose were taken from Ref.~\cite{Bazela1976PSSA}. The lattice parameters for the orth and hex structures are related as $a_{\rm orth} = c_{\rm hex}$, $b_{\rm orth} = a_{\rm hex}$, $c_{\rm orth} = \sqrt 3a_{\rm hex}$, and $V_{\rm orth} = 2V_{\rm hex}$\cite{Johnson}. As one can see in Fig.~\ref{Fig1}(b), at $T_t \simeq 95$~K, $a_{\rm orth}$ and $c_{\rm orth}$ show an abrupt increase of $\sim 12$\% and $\sim 0.15$\% from the $c_{\rm hex}$ and $\sqrt{3}a_{\rm hex}$ values, respectively while $b_{\rm orth}$ shows a drastic decrease of $\sim 8.5$\% from the $a_{\rm hex}$ value. The unit cell volume at $T_t$ also shows a sudden change of $\Delta V/V_{\rm hex}\simeq 2.6$\% [see Fig.~\ref{Fig1}(c)], reflecting a significant atomic displacement during the structural transformation. This change in unit cell volume is consistent with the value reported for Fe doped MnNiGe (Ref.~\cite{Liu2012NatureCom}) and Al doped MnNiGe (Ref.~\cite{Quetz2013JAP})  systems. The temperature evolution of the fraction of the hex and orth phases for both cooling and heating protocols are shown in inset of Fig.~\ref{Fig1}(a). The high temperature hex phase fraction decreases while the orth phase fraction increases below $T_{\rm t}$. Our lattice constants, unit cell volume,  and phase fractions during cooling and heating show a thermal hysteresis ($\Delta T \simeq 10$~K) at $T_t \simeq 95$~K confirming first order nature of the transition. At 23~K, the orth and hex phase fractions are estimated to be $\sim 60$\% and $\sim 40$\%, respectively.
\begin{figure}[t!]
	\centering
	\includegraphics[width=\linewidth,height=\linewidth]{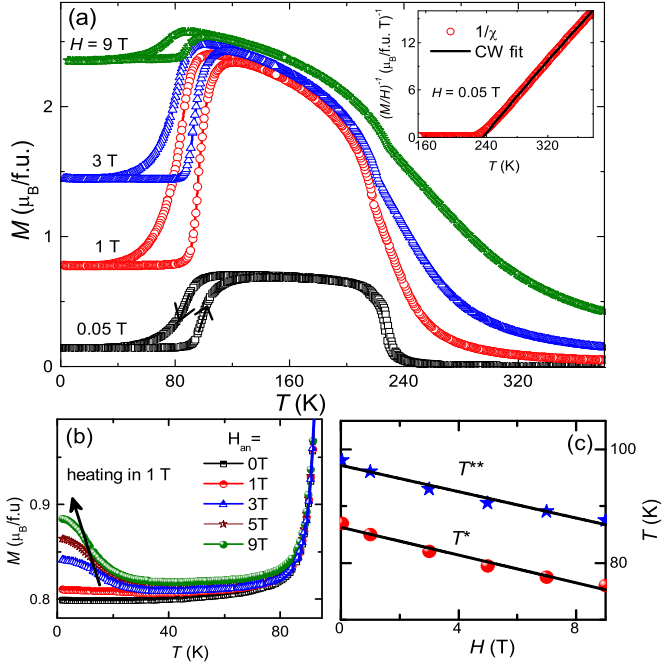}
	\caption{(a) $M(T)$ of MnNiGe$_{0.9}$Ga$_{0.1}$ measured in different applied fields during both cooling and heating. Inset: $(M/H)^{-1}$ vs. $T$ for $H=0.05$~T. (b) $M(T)$ measured in $H=1$~T during heating after the sample is cooled in different applied magnetic fields ($H_{\rm an}$). (c) Variation of $T^{*}$ and $T^{**}$ with $H$ for cooling and heating protocols.}
	\label{Fig2}
\end{figure}

Figure~\ref{Fig2}(a) shows the temperature dependent magnetization, $M(T)$ of MnNiGe$_{0.9}$Ga$_{0.1}$ measured in different applied fields during cooling and heating. With decreasing temperature, $M (T)$ data at $H=0.05$~T show an abrupt increase at around 226~K and then attains a plateau. With further decrease in temperature, $M(T)$ decreases rapidly at low temperatures. All these features point towards a PM to FM transition at  $T_C\simeq 226$~K and a FM to AFM transition at low temperatures\cite{Shi2013PSSA,Zhang2013JAP,Dutta2016JPDAP}. The FM-AFM transition temperature is estimated to be $T^*\simeq 87$~K and $T^{**}\simeq 98$~K for the cooling and heating protocols, respectively. Thermal hysteresis between the cooling and heating curves implies a first order magnetic transition at low temperatures. The average value of the transition temperature ($\sim93$~K) coincides with $T_t$ suggesting that the FM-AFM and structural transitions occur simultaneously at $T_t$. Inset of Fig.~\ref{Fig2}(a) shows $(M/H)^{-1}$ vs. $T$ for $H = 0.05$~T. A fit of the high-$T$ data ($\geq 275$~K) by Curie-Weiss law,
\begin{equation}
\dfrac{M}{H}=\chi_0 + \dfrac{C}{(T-\theta_{\rm p})}
\end{equation}
where $\chi_0$ is the temperature independent susceptibility and $C$ is the Curie constant. It gives an effective moment, $\mu_{\rm {eff}} \simeq 6.1~\mu_{\rm B}$/f.u. and $\theta_{\rm p} \simeq 235$~K for the high temperature Ni$_2$In-type hex phase. Since our $T_t$ is much smaller than $\theta _{\rm p}$ of the Ni$_2$In-type hex phase, a strong coupling between magnetic and structural transitions is in accordance with the previous predictions\cite{Liu2012NatureCom,Zhang2008APL,zhang2010JPDAP}. It is reported that the magneto-structural transition $T_t$ moves towards low temperatures linearly with increasing Ga concentration\cite{Zhang2013JAP,Shi2013PSSA}. Our observed value of $T_t$ for 10\% Ga indeed falls on the same line.

Well below $T_{\rm t}$, a non-zero value of $M$ for $H=0.05$~T suggests that a finite fraction of the Ni$_2$In-type hex FM phase exists in the low temperature TiNiSi-type orth AFM state\cite{Zhang2013JAP,Shi2013PSSA}. This is indeed consistent with the XRD results where we have quantitatively estimated the phase fractions. As field increases, the value of $M$ below $T_{\rm t}$ increases drastically suggesting a field induced AFM to FM transition in this alloy. Dutta et. al.\cite{Dutta2016JPDAP} also found similar behaviour in the presence of external pressure. They speculated that the presence of some of the arrested FM phase fraction along with the AFM phase could also be responsible for such an increase in the overall $M$ at low temperatures. In order to ascertain this, $M(T)$ was measured under cooling and heating in unequal magnetic field protocol\cite{Banerjee2006JPCM} i.e. $M(T)$ was measured during heating under $H=1$~T after cooling at different fields ($H_{\rm an}$) [see Fig.~\ref{Fig2}(b)]. For $H_{\rm an}>1$~T (e.g. at $H_{\rm an} = 3$~T), $M(T)$ shows a clear upturn below about $T_{\rm g} \sim 14$~K. As $H_{\rm an}$ increases, $T_{\rm g}$ is found to shift towards lower temperatures whereas $T^{**}$ is not at all affected. This type of feature suggests the transformation of a finite fraction of the low temperature FM phase to AFM phase at $T_g$ followed by a AFM to FM transition at $T^{**}$. This is reminiscent of a re-entrant (FM-AFM-FM) transition, akin to that observed for Pd doped FeRh (Ref.~\cite{Kushwaha2009PRB}), Co doped Mn$_2$Sb (Ref.~\cite{Kushwaha2007JPCM}), Ru doped CeFe$_2$ (Ref.~\cite{Roy2004PRL}) etc and is described to be due to kinetic arrest of the first order magnetic transition. It also suggests that the high field FM phase is in the non-equilibrium state and the low field AFM phase is in the equilibrium state. As shown in Fig.~\ref{Fig2}(c), for both heating and cooling protocols, FM-AFM transition is found to be varying linearly with field with a slope of $\sim 1.2$~K/T.

\begin{figure}[t!]
	\centering
	\includegraphics[width=\linewidth,height=.6\linewidth]{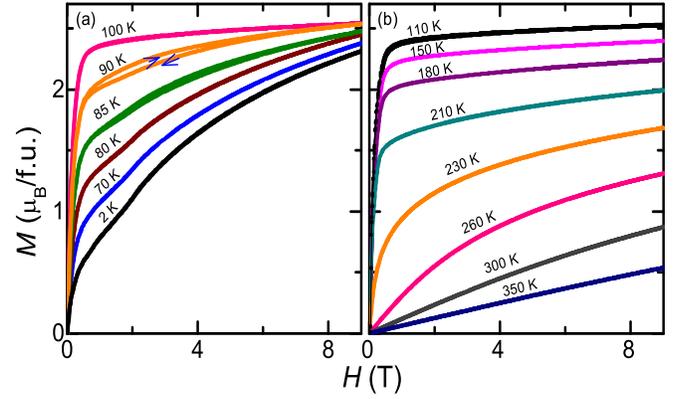}
	\caption{Magnetic isotherms measured at different temperatures (a) around $T_t$ and (b) around $T_{\rm C}$.}
	\label{Fig3}
\end{figure}
Magnetic isotherms [$M(H)$] measured at different temperatures are shown in Fig.~\ref{Fig3}. For each temperature, the measurements were performed after the zero-field-cooling from 300~K. At $T = 350$~K, it is nearly a straight line, as expected in the PM regime. With decreasing temperature, it develops a curvature and saturates completely at around 110~K reflecting a purely FM phase in this temperature regime, consistent with our $M(T)$ measurements. The spontaneous magnetization at 110~K was estimated from the $y$-intercept of the linear extrapolation of the high field data to be $\sim 2.4~\mu_B$/f.u. which is consistent with the magnetic moment of Mn ($\sim 2.3~\mu_B$/f.u.) in the parent MnNiGe compound\cite{Fjellvaag1985JMMM}. At $T \simeq 90$~K, a magnetic hysteresis was observed during the process of increasing and decreasing fields. For lower temperatures ($T \leq T_t$), the shape of the $M(H)$ curve changes drastically. First, $M$ increases rapidly upto $H\simeq 0.5$~T and the increase is slowed down for higher fields. The sharp increase of $M$ in the low field regime indicates the presence of a finite fraction of the FM phase which is saturated for $H\leq 0.5$~T. With decreasing temperature, the value of $M$ at $H\simeq 0.5$~T decreases which indicates a decrease of FM phase fraction with temperature. For $H> 0.5$~T, the gradual increase of $M$ and the lack of complete saturation even at 9~T clearly reflects the presence of majority AFM phase.

\begin{figure}[t!]
	\centering
	\includegraphics[width=\linewidth,height=\linewidth]{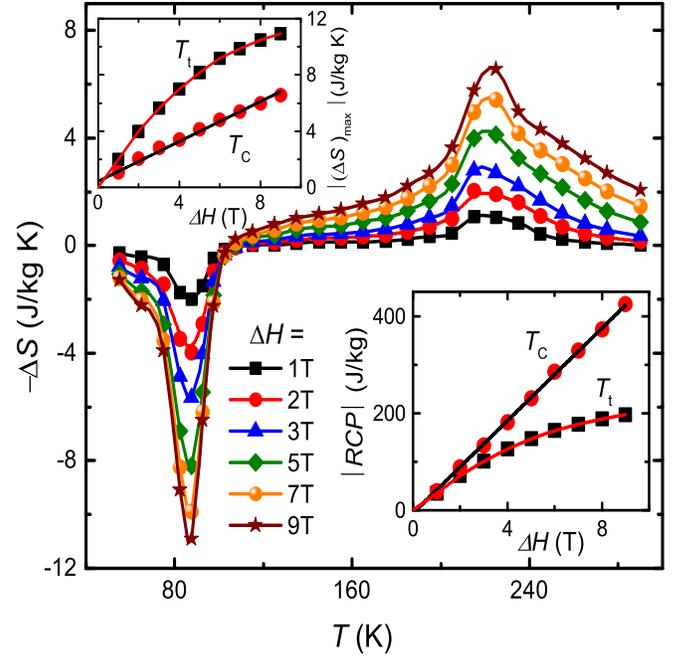}
	\caption{Temperature dependent isothermal change in magnetic entropy ($\Delta S$) for MnNiGe$_{0.9}$Ga$_{0.1}$ under different field variation ($\Delta H$). Upper and lower insets show the variation of $(\Delta S)_{\rm max}$ and $RCP$ with $H$ across $T_t$ and $T_C$, respectively.}
	\label{Fig4}
\end{figure}

To make an assessment of the MCE in this alloy, we estimated the isothermal change in entropy ($\Delta S$). Figure~\ref{Fig4} displays $\Delta S$ as a function of temperature for different changes in field ($\Delta H$) derived from the $M (H)$ data (in Fig.~\ref{Fig3}) using the Maxwell relation\cite{Li2012EPL}, 
\begin{eqnarray}
& &\Delta S(T_{av}, H) = \int_{0}^{H} \dfrac{\partial M}{\partial T}dH \nonumber \\
&\approx& \dfrac{1}{T_{i+1}-T_i}\int_{0}^{H}[M(T_{i+1},H)-M(T_i,H)]dH.\nonumber\\
 & &
\end{eqnarray}
Here, $T_i$ and $T_{i+1}$ are the two successive temperatures at which $M(H)$s were measured and $T_{av}$ is the average of $T_i$ and $T_{i+1}$. A large inverse MCE was observed in the vicinity of $T_t$ due to an abrupt change of $M$ and a conventional MCE was observed near $T_C$. The maximum value of $\Delta S$ [$(\Delta S)_{\rm max}$] near $T_t$ is found to be $\sim 4$~J/kg~K, $\sim 8.2$~J/kg~K, $\sim 9.9$~J/kg~K, and $\sim 10.9$~J/kg~K for $\Delta H=2$~T, 5~T, 7~T, and 9~T, respectively, which varies almost exponentially with $\Delta H$ (upper inset of Fig.~\ref{Fig4}) and is not saturated even upto 9~T. On the other hand, our estimated value of $(\Delta S)_{\rm max}$ at $T_C$ increases almost linearly with $\Delta H$ (upper inset of Fig.~\ref{Fig4}). The values $\sim-2$~J/kg~K, $\sim-4.1$~J/kg~K, $\sim-5.4$~J/kg~K, and $\sim-6.5$~J/kg~K for $\Delta H=2$~T, 5~T, 7~T, and 9~T, respectively, are consistent with other reported systems\cite{Dutta2016JPDAP,Shi2013PSSA,Quetz2013JAP,Samanta2012APL}.

It is predicted that as $T_{\rm t}$ moves away from $T_{\rm C}$ either due to chemical or hydrostatic pressure, the change in entropy across $T_{\rm t}$ decreases\cite{Gottschall2016PRB}.  For the series MnNiGe$_{1-x}$Ga$_x$, MCE has been studied for $x=0.07$ to 0.085. Indeed, $T_{\rm t}$ is found to be moving towards lower temperatures with increasing $x$ and without affecting the $T_{\rm C}$\cite{Zhang2013JAP,Shi2013PSSA,Dutta2016JPDAP}. For our system ($x=0.1$), the value of $(\Delta S)_{\rm max}$ for $\Delta H=2$~T and 5~T is found to be larger than that reported for $x=0.072$ at $T_t \simeq 150$~K by Dutta et al\cite{Dutta2016JPDAP} and smaller than that reported by Shi et al\cite{Shi2013PSSA} for $x=0.0725$, 0.080, and 0.085 at $T_t\simeq 202$~K, 175~K and 143~K, respectively. Thus, for the nearly same composition, different values of $(\Delta S)_{\rm max}$ are reported in different literatures suggesting that the $(\Delta S)_{\rm max}$ value is very much sample dependent. It is also to be noted that if the magneto-structural transition at $T_{\rm t}$ is complete then one should expect a higher value of $(\Delta S)_{\rm max}$. However, all the compounds in the series MnNiGe$_{1-x}$Ga$_x$ seem to have in-complete magneto-structural transitions and co-existing phases below $T_{\rm t}$. The ratio of co-existing phases below $T_{\rm t}$ can vary with $x$ which may affects the value of $(\Delta S)_{\rm max}$ significantly at $T_{\rm t}$. However, a clear understanding on this issue requires the estimation of phase fractions (as reported here for $x=0.1$) below $T_{\rm t}$ for a series of $x$ values synthesised under same conditions.
 
It worth mentioning that our $(\Delta S)_{\rm max}$ at $\Delta H=5$~T for $x=0.1$ is also comparable to other well known MCE materials such as Mn$_2$NiGa,\cite{Singh2014APL} Ni$_{50}$Mn$_{36}$Sb$_{14}$,\cite{Du2007JPDAP} Ni$_{50}$Mn$_{34}$In$_{16}$,\cite{Aksoy2007APL} Mn$_5$Ge$_3$,\cite{Zhang2007JAP} and Gd$_5$Ge$_{1.9}$Si$_2$Fe$_{0.1}$\cite{Provenzano2004Nature}. It is even larger than what is reported for MnNi$_{0.895}$Cr$_{0.105}$Ge$_{1.05}$\cite{Aryal2015JAP} and MnCo$_{0.9}$Fe$_{0.1}$Si \cite{Xu2014JPDAP}.

The relative cooling power (RCP) which characterizes the performance of a magnetic refrigerant material is defined as\cite{Tishin1999Book}
\begin{equation}
RCP=(\Delta S)_{\rm max}\times \delta T_{\rm FWHM},
\end{equation} 
where $\delta T_{\rm FWHM}$ is the full width at half maximum of the peaks in the $\Delta S$ vs. $T$ curves. As shown in the lower inset of Fig.~\ref{Fig4}, the value of $RCP$ also increases exponentially with increasing $\Delta H$ at $T_t$, whereas at $T_C$ the increase is linear with $\Delta H$. It has a value $\sim 147$~J/kg~K and $\sim -230$~J/kg~K at $T_t$ and $T_C$, respectively for $\Delta H=5$~T. These values are reasonably high as compared to other MCE alloys\cite{zhang2010JPDAP,Zhang2007PRB,Dutta2016JPDAP,Pathak2010APL}. Thus, because of the large values of $\Delta S$ and RCP with an inverse MCE at $T_t$ and a conventional MCE at $T_C$, the alloy under investigation can be viewed as a potential candidate for magnetic refrigeration purpose. Another advantage of this alloy is that due to small hysteresis in the $M(H)$ curve at $T=90$~K, the energy loss is minimum ($\sim 8$~J/kg) compared to other Mn-based systems\cite{Nayak2010JAP,Stadler2006APL,Pathak2010APL}. 
\section{Conclusion}
In summary, we have studied in detail the structural and magnetic properties of MnNiGe$_{0.9}$Ge$_{0.1}$. It reveals a first order MST at $T_t \simeq 95$~K, below which both Ni$_2$In-type hex FM and TiNiSi-type orth AFM phases coexist. At 23~K, fraction of the FM and AFM phases is estimated to be $\sim 40\%$ and $\sim 60\%$, respectively. The change in unit cell volume at $T_t$ between the two phases is found to be $\sim 2.6\%$. Further, the coexisting phases give rise to a re-entrant transition when the sample is cooled in different applied fields which suggest that the low temperature FM phase is non-equilibrium state. The coupling of magnetic and structural transitions leads to a large inverse MCE at $T_t$ and a conventional MCE at $T_C$, suggesting that this alloy can be considered as a potential material for magneto-refrigeration.

\acknowledgments
We acknowledge R. Rawat for his useful suggestions. PB was supported by the IISER-TVM, Post-doctoral programme.
 \bibliographystyle{eplbib} 
\providecommand{\noopsort}[1]{}\providecommand{\singleletter}[1]{#1}%

\end{document}